\documentclass{article}

\usepackage[utf8]{inputenc}
\usepackage{microtype}
\usepackage[numbers]{natbib}
\usepackage[toc,page]{appendix}
\usepackage{fullpage}
\usepackage{times}
\usepackage{fancyhdr,graphicx,amsmath,amssymb}
\usepackage{algorithm}
\usepackage{algorithmic}
\usepackage{multirow}
\usepackage{rotating}
\usepackage{tablefootnote}
\usepackage{xcolor}
\usepackage{url}

\begin{document} 

\begin{center}
    \textbf{Predicting Elite NBA Lineups Using Individual Player Order Statistics}
    
    Susan E. Martonosi$^{1}$, Martin Gonzalez$^{1}$, Nicolas Oshiro$^{2}$
    
    1: Harvey Mudd College
    
    2: University of San Francisco
\end{center}

\textbf{Abstract: }
NBA team managers and owners try to acquire high-performing players.  An important consideration in these decisions is how well the new players will perform in combination with their teammates.  Our objective is to identify elite five-person lineups, which we define as those having a positive plus-minus per minute (PMM). Using individual player order statistics, our model can identify an elite lineup even if the five players in the lineup have never played together, which can inform player acquisition decisions, salary negotiations, and real-time coaching decisions. We combine seven classification tools into a unanimous consent classifier (all-or-nothing classifier, or ANC) in which a lineup is predicted to be elite only if all seven classifiers predict it to be elite. In this way, we achieve high positive predictive value (i.e., precision), the likelihood that a lineup classified as elite will indeed have a positive PMM.  We train and test the model on individual player and lineup data from the 2017-18 season and use the model to predict the performance of lineups drawn from all 30 NBA teams' 2018-19 regular season rosters.    Although the ANC is conservative and misses some high-performing lineups, it achieves high precision and recommends positionally balanced lineups. \textbf{Keywords:} Basketball; point differential; lineups; classification.

\section{Introduction}
\label{s:intro}

The strategy of professional basketball continues to evolve, leading NBA team managers and coaches to continuously seek new players who have the relevant skills to win the annual championship.  While individual player performance is a significant contributor towards team success, we are interested in the overarching factors contributing to  high-performing teams and the collection of players that comprise them.  This leads us to develop a model to predict high-performing five-person lineups, including those which have never played together before, using individual player order statistics. This model can help general managers considering trades and free agents evaluate how potential incoming players might perform in combination with existing team members. Coaches can also use this model as an in-game aid to see which player sitting on the bench may have a positive impact as a substitute.

Much of the work in the literature focuses on methods for predicting the outcomes of basketball games, tournaments, or seasons. \"{O}zmen quantifies the incremental change in win probability per unit change in several game statistics \cite{ozmen}.  Loeffelholz \textit{et al.} use neural networks to predict the outcomes of NBA games more successfully than expert prediction \cite{Loeffelholz}.  Lin uses historical team data with real-time updating during a game to predict the game's outcome \cite{Lin}. Shen \textit{et al.} and Hua present methods for predicting the results of the NCAA ``March Madness'' collegiate tournament \cite{ShenEtAl, Hua}, as do several articles in a 2015 special issue of this journal \cite{GlickmanSonas}.  Gumm \textit{et al.} and Zimmerman \textit{et al.} 
use machine learning approaches on historical data to predict outcomes of NCAA and NBA playoffs \cite{GummEtAl, Zimmermann2013PredictingNM}.   
Vaz de Melo \textit{et al.} forecast NBA outcomes using network analysis \cite{VazdeMeloEtAl}, while Cheng \textit{et al.} use the maximum entropy principle to predict the outcome of NBA playoffs \cite{ChengEtAl}. Ruiz and Cruz present a model to predict the win probability of teams participating in the NCAA March Madness Tournament by combining a Poisson factorization method with a model borrowed from soccer which quantifies a team's attack and defense coefficients \cite{ruiz2015generative}. 

A body of work examines the role of individual players in determining team success.  Several papers use centrality metrics and other methods from the field of complex social networks to identify prominent players and evaluate team performance \cite{ClementeEtAl, RibeiroEtAl, WascheEtAl}.  Deshpande and Jensen quantify an individual player's contribution to their team's overall win probability \cite{DeshpandeJensen}. However, it has also been suggested that an individual's contribution towards a team's success, as measured by real Plus-Minus, cannot be disentangled from the influence of other players \cite{ghimire2020measuring}. While statistics such as Box Plus Minus (BPM) attempt to measure an individual player's contributions to the team relative to that of their teammates, BPM does not give an indication of how a five-player lineup will perform together.

Our paper builds upon work that examines the effectiveness of five-person lineups in the NBA. For instance, Maymin \textit{et al.} calculate the synergies of each NBA team by comparing their 5-player lineups’ effectiveness to the ``sum-of-the-parts'' \cite{maymin}. Robertson uses a graphical model to capture the series of events that occur during a possession and estimate the probabilities of game-play events given the players on the floor \cite{robertson}.  
Kalman and Bosch redefine traditional basketball positions using model-based clustering and identify how interactions between these new positions result in lineups with the highest net-rating \cite{kalman2020nba}. Pelechrinis presents LinNet, which exploits the dynamics of a directed network that captures the performance of lineups against a specific opponent lineup \cite{pelechrinis}. In addition, Sisneros and Van Moer use point differential (plus-minus) to measure the contribution of individual players towards a team’s success, and predicts the winning percentage of a given team \cite{sisnerosandvanmoer}. Oh \textit{et al.} simulate game play using a probabilistic network model; however, a limitation of their work is the inability to predict the performance of a lineup using players from outside the team \cite{OhEtAl}.

Our work contributes towards this general discussion by developing a model that can predict the performance of unseen lineups, as measured by the predicted sign of their point differential per minute. While most of the work we have encountered focuses on guiding coaching decisions for in-game substitutions, our work goes beyond and can be used as a tool for free agency decisions, trade negotiations, and calling up players from the NBA G league. 

Our model identifies lineups with a high probability of contributing a positive point differential, also known as plus-minus, per minute (PMM). A positive PMM indicates that the lineup generally contributes positively to the team’s relative score, whereas a negative PMM indicates that the lineup generally detracts from the team’s relative score. Our objective is to predict elite lineups, which we define to be those having a high likelihood of contributing a positive PMM.  We combine seven classification tools into a unanimous consent classifier, which we call the \textit{all-or-nothing classifier} (ANC).  The ANC predicts a lineup to be elite only if all seven subclassifiers predict it to be elite. In this way, we achieve high \textit{precision}, the likelihood that a lineup classified as elite will indeed have a high PMM.  Each tool takes as input the individual player order statistics for the five players on a proposed lineup, capturing the lineup's offensive and defensive capabilities, and classifies the lineup as either elite or not.  The subclassifiers used are a decision tree, random forest, boosting, a support vector machine, k-nearest neighbors, logistic regression, and linear discriminant analysis.

We train and test the model on individual player data from the 2017-18 NBA regular season purchased from BigDataBall.com and supplemented with hustle stats from NBA.com. Hustle stats are included since they quantify a player’s defensive contribution to the lineup. To train the models, we use a random sample of 712 lineups with at least 25 minutes of playing time.  We then test the model on a random sample of 176 lineups having at least 25 minutes of playing time.  We achieve a precision of 86.7\% on the testing set, indicating that lineups classified as elite using the ANC are highly likely to contribute a positive PMM to the team.   We then train the model on the full 2017-18 NBA regular season data and use the trained model to predict elite lineups for all 30 NBA teams 
during the 2018-19 regular season.  We find that the classifier achieves a  high precision of 76.9\% (compared to 62.1\% prevalence of elite lineups), even when used to make predictions from one season to the next.

This paper will be structured as follows. In Section \ref{s:methodology}, we provide an overview of the methodology used in our classification framework.  Section \ref{s:data} describes the datasets we use to train and test our model and outlines the steps taken to clean and merge the data.  Section \ref{s:results} presents results of applying our model to a test data set as well as to making predictions for the 2018-19 regular season,  
and we compare our predictions against actual season outcomes.   We present suggestions for future work and our conclusions 
in Section \ref{s:conclusion}.  
 
\section{Methodology}
\label{s:methodology}

In this section, we describe the classification problem our method solves, the \textit{all-or-nothing classifier} (ANC) we develop to solve it, and the method of cross-validation we use to tune hyperparameters to achieve a reliably high positive predictive value, or precision.

\subsection{Classifying the Plus-Minus Per Minute of an Unseen Lineup}
\label{ss:classifypmm}

The performance of a five-person lineup is often summarized in their plus-minus per minute (PMM) metric.  This measures the cumulative point differential (points earned minus points scored by the opponent) accrued by the lineup each time they appear together on the court, divided by their total playing time together as a lineup.  Lineups with positive PMM contribute to the team’s net score per unit time, while lineups with negative PMM give up more points than they earn per unit time.  Existing methods for predicting the performance of unseen lineups, such as the adjusted plus-minus, attempt to account for a player's individual contribution to the score after controlling for the other players on the court \cite{winston}.  However, these methods are based solely on fitting the observed plus-minus to indicators of who is or is not on the court; they do not incorporate other statistics about individual player performance.

The method we pursue is to predict an unseen lineup’s PMM using individual player statistics.  For a useful primer on basketball statistics, see Kubatko \textit{et al.} \cite{KubatkoEtAl}.  Let $L$ be the set of all five-person lineups, and let $P$ be the set of players.  For each lineup $l \in L$, let $p(l) \in P$ be the set of five players included in lineup $l$.  For each player $p \in P$, let $s_{p1}, \ldots, s_{pn}$ be the collection of statistics observed on a per-minute basis, e.g., field goals per minute played, or defensive rebounds per minute.  We thus wish to predict the plus-minus per minute of lineup $l$, denoted $PMM(l)$ as a function of the values of $s_{p1}, \ldots, s_{pn}$  for each player $p \in  p(l)$:
$$PMM(l) = f(s_{pi}| p \in  p(l), i \in  1, \ldots , n).$$

Earlier work demonstrates the challenge of predicting $PMM(l)$ directly \cite{bynum}.  Thus, in this work we focus on predicting the sign of $PMM(l)$, classifying a lineup $l$ as \texttt{elite} if $PMM(l) > 0$.

We make the assumption that the connection between PMM and player statistics is distributional. For each lineup and each statistic, we sort the values of the statistic from smallest to largest for the five players on that lineup.  We then use these order statistics as predictors in the model.  This is effectively predicting $sign(PMM)$ using the $20^{th}, 40^{th}, 60^{th},  80^{th}$, and $100^{th}$ quantiles of the distributions of each statistic for the five players on the lineup.  For instance, one player might contribute highly to field goal percentage, while another player is contributing highly to blocks, and all of these factors contribute to the overall PMM experienced by the lineup.

As a small example, consider the two player statistics of field goals per minute (FGM) and defensive blocks per minute (DBM) in the context of Lineup 1 (Jaylen Brown, Kyrie Irving, Marcus Morris, Terry Rozier, Jayson Tatum of the Boston Celtics) and Lineup 2 (James Harden, Chris Paul, Eric Gordon, Nene Hilario, Clint Capela of the Houston Rockets), during the 2017-18 regular season.  We list each player’s values for the two statistics in Tables \ref{tab:Celtics} and \ref{tab:Rockets} \cite{bbrefCeltics, bbrefRockets}.

\begin{table}
\caption{Example individual player statistics for Lineup 1 \cite{bbrefCeltics}.}
\label{tab:Celtics}
\begin{tabular}{llllll}

 & Jaylen  & Kyrie  & Marcus  &  Terry  &	Jayson \\ 
Player &  Brown &  Irving &  Morris &   Rozier &	 Tatum\\ \hline
 Field Goals per Minute	& 0.17	& 0.28	& 0.18 &	0.15 &	0.16 \\
Defensive Blocks per Minute	& 0.01	& 0.01	& 0.01	& 0.01	& 0.02 \\
\end{tabular}
\end{table}

\begin{table}
\caption{Example individual player statistics for Lineup 2 \cite{bbrefRockets}.}
\label{tab:Rockets}
\begin{tabular}{llllll}

	& James 	& Chris 	& Eric 	& Nene 	& Clint \\ 
Player	&  Harden	&  Paul	&  Gordon	&  Hilario	&  Capela\\ \hline
Field Goals per Minute &	0.26 &	0.20 &	0.19 &	0.18 &	0.22 \\
Defensive Blocks per Minute	& 0.02	& 0.01 &	0.01 &	0.02 &	0.07 \\
\end{tabular}
\end{table}

For each lineup, we sort the five values of FGM from smallest to largest and the five values of DBM from smallest to largest to obtain the ten predictors $FGM_{(1)}$, $FGM_{(2)}$, $\ldots$, $FGM_{(5)}$, $DBM_{(1)}$, $DBM_{(2)}$, $\ldots$, $DBM_{(5)}$, where the subscript $(i)$ represents the $i^{th}$ smallest value of the five players (known as the $i^{th}$ order statistic).  For the two lineups given, this results in the two rows of the data set shown in Table \ref{tab:expredictors}. Thus, our set of predictors is the order statistics of the five players on the lineup, for each player performance statistic collected. By using  the order statistics for both offensive and defensive metrics, we can capture the overall playing profile of the lineup. 

\begin{sidewaystable}
\caption{Example player order statistics for the two lineups given in Tables \ref{tab:Celtics} and \ref{tab:Rockets}.  These order statistics serve as ten predictors in the classification framework.}
\label{tab:expredictors}
\begin{tabular}{lllllllllll}

Lineup	& $FGM_{(1)}$& 	$FGM_{(2)}$	& $FGM_{(3)}$& 	$FGM_{(4)}$& 	$FGM_{(5)}$& 	$DBM_{(1)}$& 	$DBM_{(2)}$& 	$DBM_{(3)}$& 	$DBM_{(4)}$& 	$DBM_{(5)}$ \\ \hline
1	& 0.15	& 0.16	& 0.17	& 0.18	& 0.28	& 0.01	& 0.01	& 0.01	& 0.01	& 0.02 \\
2	& 0.18	& 0.19	& 0.20	& 0.22	& 0.26	& 0.01	& 0.01	& 0.02	& 0.02	& 0.07 \\
\end{tabular}
\end{sidewaystable}

\subsection{Choice of Classifier}
\label{ss:classchoice}

In order for our classifier to be an effective decision support tool for personnel decisions, we want to be confident that lineups flagged as \texttt{elite} by our classifier will indeed perform well.  Most classification frameworks strive to maximize the overall accuracy of the classifier, which is the probability that the classification given by the model matches the true class of the datum.  However, in the case of identifying elite basketball lineups, we are more interested in achieving a high positive predictive value, or precision. This is the conditional probability that a lineup is actually elite (has a positive PMM) given that the classifier predicts it to be elite.  

\begin{figure}
\includegraphics[width =\textwidth]{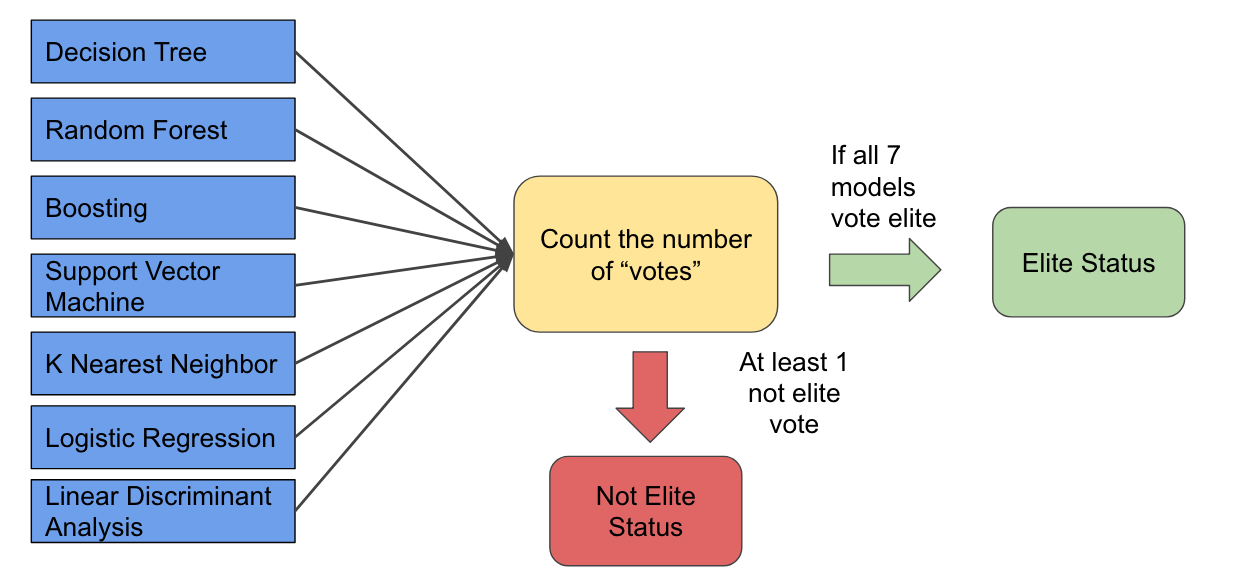}
\caption{The decision process used by the all-or-nothing classifier to categorize lineups.}
\label{graph:decproc}
\end{figure}

To enhance the precision of the classifier, we combine seven commonly used classifiers (which we will refer to as subclassifiers) into an \textit{all-or-nothing classifier} (ANC), in which a lineup is classified as \texttt{elite} if and only if all seven subclassifiers predict it to be so.  Consensus classifiers have successfully been used in other applications, including computational biology \cite{BendlEtAl}.  The seven subclassifiers are a decision tree, a random forest, boosting, a support vector machine, $k$-nearest neighbors, logistic regression, and linear discriminant analysis.  Each of these algorithms casts a vote on the classification of a given lineup based on its individual player order statistics. A schematic of the ANC is shown in Figure \ref{graph:decproc}. 
 
\subsection{Parameter Tuning}
\label{ss:paramtune}

Each of these subclassification methods has a set of parameters governing the algorithm.  Rather than tuning each subclassifier individually, we tune the parameters as an ensemble to achieve a reliably high precision of the overall ANC classifier. We use $10$-fold cross validation on a training set of lineups. (The data are described in greater detail in Section \ref{s:data}.)  The parameters associated with each subclassifier and with the overall ANC classifier are described briefly here.  Then we provide pseudocode outlining the ensemble tuning procedure.

\subsubsection{Decision Tree Parameters}
\label{sss:dectreparams}

A decision tree iteratively determines threshold values of the predictor variables along which to split the classification of a lineup as \texttt{elite} or not.  Common practice is to build a full tree and then prune it according to a cost-complexity tradeoff.  The cost-complexity parameter \texttt{cp} in R’s \texttt{rpart} package for building and pruning decision trees represents the percentage reduction in inter-branch variability required to justify each subsequent split.  We use $cp=-1$ to build the initial complete tree, and then tune the value of \texttt{cp} used to prune the tree. Additionally, we use the \texttt{weights} argument to weight the observations by the total number of minutes played by each lineup so that lineups having low total playing time have less influence on the fitting of the tree. To prioritize achieving a high positive predictive value, we tune the false positive element of the \texttt{loss} matrix option in \texttt{parms} to more heavily penalize misclassifying a \texttt{not elite} lineup as \texttt{elite}.  All other parameters are assigned the default value given in the \texttt{rpart} function of R.

\subsubsection{Random Forest Parameters}
\label{sss:rfparams}

In a random forest classifier, many decision trees are built on bootstrapped samples of the training data, and the subset of predictors considered in each splitting decision is randomly selected from the full set of predictors.  The classifications yielded by each tree are used as votes in the forest classification of an individual observation.  The parameter that governs this voting in a binary classifier is the cutoff vector $(c, 1-c)$. If we observe a fraction $p$ of trees classifying a lineup as \texttt{elite} and a fraction $1-p$ of trees classifying the lineup as \texttt{not elite}, the forest classification would be the category achieving the maximum of $(p/c, (1-p)/(1-c))$.  In other words, $c$ represents the proportion of trees yielding an \texttt{elite} classification at which the forest classifier would be ambivalent between the two groups.    Because we seek to maximize the precision of the ANC, we tune our random forest using values of $c \geq 0.5$. We also tune the parameter \texttt{ntree} governing the number of trees in the forest. All other parameters are assigned the default value given in the \texttt{randomForest} function in R’s \texttt{randomForest} package, including the parameter \texttt{mtry}, governing the number of predictors included in each subset, whose default value is the square root of the number of predictors.  

\subsubsection{Boosting Parameters}
\label{sss:boostparams}

The boosting algorithm iteratively builds a classification tree such that subsequent splits are themselves determined by decision trees fit to the residuals (misclassifications) from earlier splits. The three parameters tuned in our classifier are the number of trees to build (referred to as \texttt{mfinal} in R’s \texttt{boosting} function of the \texttt{adabag} package), the depth of each tree (\texttt{maxdepth}) and the cost complexity value (\texttt{cp}) used to prune each tree.  Additionally, we weight the observations by the total number of minutes played by each lineup to de-emphasize lineups with low playing times. All other parameters are assigned the default value given in the \texttt{boosting} function of the \texttt{adabag} package.   

\subsubsection{Support Vector Machine Parameters}
\label{sss:svmparams}

Our support vector machine (SVM) subclassifier uses a radial basis kernel and tunes two parameters, \texttt{cost} and \texttt{gamma}:  \texttt{cost} represents the penalty associated with misclassification, and \texttt{gamma} controls the rate of decay of influence an individual support vector has on the classification of points a given distance away.  Smaller \texttt{gamma} reduces the rate of decay, indicating that the support vector can influence the classification of points farther away; larger \texttt{gamma} has the opposite effect.  Other tunable parameters are set to the default used in the \texttt{svm} function of R’s \texttt{e1071} package.

\subsubsection{$K$-Nearest Neighbors Parameters}
\label{sss:knnparams}

The $k$-nearest neighbors subclassifier labels a lineup according to the most common label of its $k$ closest lineups in the predictor space.  Thus, we tune the value of $k$, using the \texttt{knn} function of package \texttt{class}.

\subsubsection{Logistic Regression Parameters}
\label{sss:logregparams}

Logistic regression predicts the log odds, $LO$, of a lineup being elite as a linear function of the predictors.  We fit the model using the function \texttt{glm} in the R package, \texttt{stats}.  Then the classification of the lineup is determined by calculating the estimated probability of elite classification, $(1+e^{-LO})^{-1}$  and deeming a lineup \texttt{elite} if the estimated probability of being elite exceeds a stated threshold.  A natural default threshold is $50\%$; however, given our interest in maximizing precision, we tune the threshold for values greater than or equal to $50\%$, essentially requiring stronger evidence before classifying a lineup as \texttt{elite}. The parameter \texttt{thresh} in \texttt{glm} refers to one minus the desired probability threshold.

\subsubsection{Linear Discriminant Analysis Parameters}
\label{sss:ldaparams}

Linear discriminant analysis, as implemented in the \texttt{lda} function in the \texttt{MASS} package of R, does not rely on tuning parameters. However, it does permit us to weight the observations by the total number of minutes played by each lineup.

\subsubsection{ANC Parameters}
\label{sss:ancparams}

The tunable parameter for the overall ANC classifier is the number of subclassifiers identifying a lineup as \texttt{elite} required for the ANC to classify the lineup as \texttt{elite}. We refer to this as \texttt{numVotes}.  While we hypothesize that we will get the highest precision by requiring all seven classifiers to agree (as the name ``all-or-nothing classifier'' suggests), we verify this assumption by tuning the number of votes required before a lineup is classified as \texttt{elite}. 

\subsubsection{Ensemble Tuning Procedure}
\label{sss:ensembletune}

Rather than tune each subclassifier individually, we wish to select the best combination of parameters across all subclassifiers that yields a reliably high precision of the ANC in $10$-fold cross validation.  We perform grid search over the combination of parameters for all seven subclassifiers and the \texttt{numVotes} parameter.  For each parameter combination, we compute the ANC classification on each fold and calculate the precision achieved on that fold.  We then average the precision over the folds for each parameter combination.  This gives us a good metric for the expected precision of the ANC for each combination of tuning parameters.  However, in addition to maximizing average precision, we also seek a classifier that performs robustly.  That is, of those parameter combinations achieving a relatively high precision, we will favor those for which the standard deviation of precision across the $k$ folds is low (i.e., it is a robust classifier), and for which the precision on any one fold is not too low (i.e., it is a reliable classifier). We can then examine these metrics across all combinations of parameters and select parameter values that yield a sufficiently high average precision along with a high minimum precision and a low standard deviation. The pseudocode for this procedure is outlined in Algorithm \ref{alg:pseudocode}.

\begin{algorithm}[H]
\caption{Pseudocode for parameter tuning in the ANC.}
\label{alg:pseudocode}
\begin{algorithmic}
 \STATE Split the data into training and testing sets
 \STATE Split the training set into 10 folds for cross-validation
 \FOR{each fold $k$} 
    \STATE Standardize the fold-training data by subtracting the mean and dividing by the standard deviation of each numerical value over the $k-1$ folds.  Scale and shift the left-out fold by the fold-training data's means and standard deviations to prevent data leakage.
     \FOR{each subclassifier $s$} 
        \FOR{each combination $c$ of tuning parameters} 
            \STATE Fit a model to the 9 folds not in $k$.
            \STATE	Predict the classes of the observations in the $k^{th}$ fold.
            \STATE Store these predictions as \texttt{preds[k,s,c]}.
        \ENDFOR
    \ENDFOR
    \FOR{each combination $c$ of tuning parameters, including \texttt{numVotes}}
        \STATE Count the number of subclassifiers $s$, for which \texttt{preds[k,s,c]} is \texttt{elite}.  
        \STATE The ANC classification for the observations in the $k^{th}$ fold is \texttt{elite} if this number is at least as large as \texttt{numVotes}, and \texttt{not elite} otherwise.
        \STATE Compute the confusion matrix comparing the ANC classification to the known classification for the observations on the $k^{th}$ fold, and store \texttt{precision[k,c]}.
    \ENDFOR
\ENDFOR
\FOR{each combination $c$ of tuning parameters, including \texttt{numVotes}}
    \STATE Calculate the average, minimum and standard deviation of precision over the 10 folds.
\ENDFOR
\STATE Choose the combination of parameters that achieves a reliably high precision (high average value, high minimum value, low standard deviation).
\end{algorithmic}
\end{algorithm}


Because we tune all possible combinations of parameters for the seven classifiers and the ANC simultaneously, the search space grows exponentially with the number of distinct values tested for each parameter.  Thus, we use a coarse grid, selecting a few distinct values for each parameter, with values informed by preliminary testing not reported here.  The specific values tested are given in Table \ref{tab:gridvals}.

\begin{table}
\caption{Parameter values used in grid search.}
\label{tab:gridvals}
\begin{tabular}{lll}

Subclassifier	& Parameter	& Values Tested \\ \hline
Decision Tree	& \texttt{cp} (cost complexity) & 	(-1, 0.01, 0.05) \\
 & & $-1$ refers to full tree \\ 
 & \texttt{loss} (misclassification penalty) & (1, 1.5, 2) \\ \hline
Random Forest	& \texttt{c} (cutoff)	& (0.5, 0.7) \\ 
 & \texttt{ntree} (number of trees) & (100,500) \\ \hline
Boosting	& \texttt{mfinal} (number of trees) & (100,500) \\
& \texttt{maxdepth} (depth of each tree) & (1, 2, 3) \\
& \texttt{cp} (cost complexity) 	& (0.01, 0.05) \\ \hline
Support Vector Machine &	\texttt{cost} (misclassification penalty) & (0.1, 1, 10) \\
 & \texttt{gamma} (influence decay) & (0.01, 0.1, 1) \\ \hline
K-Nearest Neighbors	& \texttt{k} (number of neighbors) &	(3, 5,7) \\ \hline
Logistic Regression &	\texttt{thresh} (1-probability threshold) &	(0.05, 0.25, 0.5) \\ \hline
All-or-Nothing Classifier (ANC) &	\texttt{numVotes} (agreement required) &	(1, 2, 3, 4, 5, 6, 7) \\
\end{tabular}
\end{table}

With the model framework in place, we now describe our data.

\section{Data and Implementation}
\label{s:data}

The novelty of this work lies in predicting lineup performance using the sorted individual statistics of the players comprising the lineup.  To do this requires merging data from two sources, cleaning the data to ensure lineups and players are matched correctly, filtering the data based on minutes played to reduce noise, and splitting the data into training and testing sets.  This process is described here.

\subsection{Data Sources}
\label{ss:datasource}

We wish to make predictions and compare those predictions to actual outcomes for the 2018-19 regular NBA season, the most recent season unaffected by the Covid-19 pandemic. We train and test our model using player and lineup statistics from the 2017-18 NBA regular season, using data from several sources.  We purchased play-by-play data from BigDataBall.com that contains, for every play in each game of the season, the ten people on the court, which play occurred, and the clock time.  From this source, we are able to recreate per minute point differentials (PMM) for each lineup.  While we could also use this play-by-play data to compute individual player statistics, we instead obtained player box and hustle statistics directly from NBA.com 
for simplicity and accuracy.  

The 28 individual player statistics used to form our 140 order statistic predictors are listed in Appendix \ref{app:preds}. We focus on statistics that offer direct measurements of play rather than statistics such as the Box Plus-Minus (BPM) which itself is an aggregation of many statistics and is therefore less interpretable.  

\subsection{Data Cleaning and Filtering}
\label{ss:dataclean}

Because we are merging data from two sources, NBA.com and BigDataBall.com, we must first clean both data sets to impose consistent player naming, for instance in the use of punctuation, nicknames or suffixes.  Then we match individual player statistics to lineups, sorting each statistic from smallest to largest among the players in the lineup; this creates our vectors of order statistics which we use as predictors.  

After matching players and lineups, any players appearing in only one of the two data sets (and the lineups in which they appear) should be discarded.  There was only one such player who is discarded in this way: Ty Lawson, who has playoff statistics appearing in the NBA.com data but no regular season information in the BigDataBall.com stints data.  Therefore, lineups including Lawson are not used for training or testing.

Next, we filter out individual players (and the lineups to which they belong) having fewer than 50 minutes of playing time during the season.  This is to ensure that the individual player statistics, when adjusted per minute, are estimated with low variance.  Of the 540 players represented in the raw data, 483 meet this playing time threshold.  Likewise, we filter out lineups, comprised of these 483 players, having fewer than 25 minutes of playing time together.  Of the over 14,000 distinct NBA lineups appearing in the 2017-18 season in which each player had at least 50 minutes of playing time, 888 have at least 25 minutes of lineup playing time. It is worth noting that some previous work recommends requiring a higher playing time threshold for lineups to improve model accuracy.  However, doing so would dramatically reduce our dataset.  For example, requiring 50 minutes of lineup playing time would yield only 374 lineups on which to train and test.  Given the large number of order statistics used as predictors, doing so runs the risk of overfitting the model.  Instead, to mitigate the influence of lineups with shorter playing time, we use lineup playing time as a weight in the regression tree, linear discriminant analysis, and boosting classifiers.  

\subsection{Data Splitting}
\label{ss:datasplit}

From these 888 lineups, we use an 80\%-20\% split to create a training set of 712 lineups and a testing set of 176 lineups.  In both sets, a lineup is given the label \texttt{elite} if its PMM is strictly positive; otherwise it is labeled as \texttt{not elite}.  Our training set includes 375 lineups labeled as \texttt{elite} and 337 lineups labeled as \texttt{not elite}; our testing set includes 95 lineups labeled as \texttt{elite} and 81 labeled as \texttt{not elite}.

\section{Results}
\label{s:results}

We now describe the results of tuning the ANC parameters and applying the trained ANC to the test data.  Once satisfied with the predictive power of our trained ANC, we interpret the tuned subclassifiers to understand which predictor variables appear most important in predicting lineup quality.  We then use the ANC to make predictions about lineup combinations for the 2018-19 NBA team rosters.

\subsection{Parameter Tuning Results}
\label{ss:tuneresults}

\begin{figure}
\includegraphics[width =\textwidth]{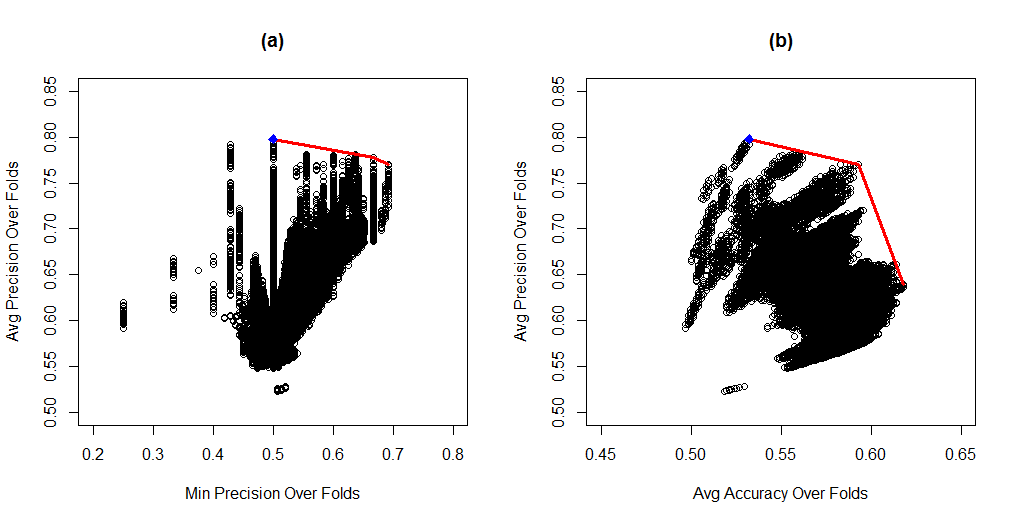}
\caption{Results of the parameter tuning.  For each combination of parameters, we plot (a) the average precision achieved by the ANC over ten folds versus the minimum precision of that classifier on the ten folds; and (b) average precision versus average accuracy.  In both plots, the efficient frontier is shown in red, and the combination of minimum and average precision achieved by our optimized parameter combination is plotted as the blue diamond.}
\label{graph:efffront}
\end{figure}

Common practice in parameter tuning is to choose the combination of parameters that achieves the best metric when averaged over the ten cross-validation folds.  In our case, that would be the parameter combination that achieves maximum average precision.  However, in our initial analysis of the parameter tuning results, we found that the combination of parameters for which the average precision is highest often has high variability in precision across the ten folds.  This implies a classifier that occasionally and unpredictably will perform poorly on certain data sets.  Figure \ref{graph:efffront}(a) shows for each parameter combination used in our tuning process the average precision achieved over ten folds against the worst-case (minimum) precision over ten folds. We exclude the results from parameter combinations in which the precision was calculated as \texttt{NA} in at least one fold.  Upon further investigation, these combinations arise exclusively when $\texttt{numVotes}=7$ and the logistic regression parameter $\texttt{thresh} = 0.05$.  In such cases, the ANC is so conservative that some folds have zero lineups classified as \texttt{elite}, yielding the value \texttt{NA} when computing precision.  Moreover, those lineups attaining a numerical value for precision have such a small number of lineups classified as \texttt{elite} that the standard deviation in precision is quite high.  We prefer a parameter combination that achieves a satisfyingly high average precision and a high worst-case precision over the ten folds. Specifically, we choose a parameter combination that lies on the efficient frontier of average precision and worst-case precision, shown in red in Figure \ref{graph:efffront}(a).  Additionally, although we are more interested in precision than overall accuracy, Figure \ref{graph:efffront}(b) plots the average precision against the average accuracy attained over ten folds in all parameter combinations, along with the efficient frontier (in red) of average precision and average accuracy.

The blue point in Figure \ref{graph:efffront}(a) corresponds to parameter combinations achieving an average precision of 79.8\% and a worst-case precision of 50\%.  Figure \ref{graph:efffront}(b) shows this point in blue on the plot of average precision versus average accuracy; we see that these parameter combinations also lie on the efficient frontier of precision and accuracy, attaining an average accuracy of 53.2\%.  We choose the optimized parameter combination to train the final model from among this set, as given in Table \ref{tab:tunedparams}.  The values of \texttt{k} (KNN), \texttt{cp} (decision tree), \texttt{loss} (decision tree), \texttt{ntree} (random forest), \texttt{c} (random forest), \texttt{thresh} (logistic regression), and \texttt{numVotes} (ANC) are uniquely determined.  However, the parameter tuning was not sensitive to the values of \texttt{mfinal} (boosting), \texttt{maxdepth} (boosting), \texttt{cp} (boosting), \texttt{cost} (SVM), or \texttt{gamma} (SVM). For these parameters, we select the values yielding the second best outcome in the efficient frontier shown in Figure \ref{graph:efffront}, corresponding to an average precision of 77.0\%, a minimum precision of 66.7\%, and an average accuracy of 59.3\%.


\begin{table}
\caption{Tuned parameter values used in final ANC.}
\label{tab:tunedparams}
\begin{tabular}{lll}

Subclassifier &	Parameter&	Chosen Value \\ \hline
Decision Tree &	\texttt{cp} (cost complexity) &	0.05 \\ 
& \texttt{loss} (misclassification penalty) & 1\\\hline
Random Forest &	\texttt{c} (cutoff) &	0.7 \\ 
& \texttt{ntree} (number of trees) & 500\\ \hline
Boosting &	\texttt{mfinal} (number of trees) & 500 \\
 &\texttt{maxdepth} (depth of each tree) & 3 \\
& \texttt{cp} (cost complexity) & 0.01 \\ \hline
Support Vector Machine	& \texttt{cost} (misclassification penalty)& 1 \\
 & \texttt{gamma} (influence decay)	& 1 \\ \hline
K-Nearest Neighbors	& \texttt{k} (number of neighbors) &	7 \\ \hline
Logistic Regression	& \texttt{thresh} ($1-$probability threshold) &	0.25 \\ \hline
All-or-Nothing Classifier (ANC) &	\texttt{numVotes} (agreement required) &	7 \\
\end{tabular}
\end{table}


We note that the performance of the classifier appears most sensitive to the number of votes required by the seven subclassifiers for a lineup to be predicted as \texttt{elite}.   Average precision begins to drop off considerably when requiring fewer than six out of the seven subclassifiers to agree, and seven is preferable to six.  Additionally, the ANC’s performance appears sensitive to the choice of the cost complexity parameter for the decision tree, the probability threshold for logistic regression, and the cutoff parameter for the random forest.  The precision of the ANC appears robust to the choice of the remaining parameters.

\subsection{Testing Results}
\label{ss:testresults}

We fit our ANC to the full, standardized, training set.  We shift and scale the testing data using the training data's means and standard deviations to avoid data leakage, and we apply our trained ANC to the testing data.  The testing data contains 176 lineups, of which 95 lineups are labeled as \texttt{elite}.    The confusion matrix is given in Table \ref{tab:confusion}.  Of 15 lineups predicted to be \texttt{elite}, 13 of these have a true label of \texttt{elite}, indicating a strictly positive PMM.  This corresponds to a precision of 86.7\%.  The model has an overall accuracy of 52.3\%, which is tolerated because our focus is on predicting elite lineups.

\begin{table}
\caption{Confusion matrix for the tuned ANC applied to the test data set. Of the 15 lineups predicted to be \texttt{elite}, 13 have a true label of \texttt{elite}, corresponding to a precision of 86.7\%.}
\label{tab:confusion}
\begin{tabular}{llll}

 && \multicolumn{2}{c}{Predicted Class} \\
& & Elite &	Not Elite \\ 
\multirow{2}{*}{\rotatebox[origin=c]{90}{\parbox[c]{1cm}{\centering True Class}}}  &  Elite &	13 &	82 \\
&	Not Elite &	2	& 79 \\
\end{tabular}
\end{table}

\begin{figure}
\includegraphics[width =\textwidth]{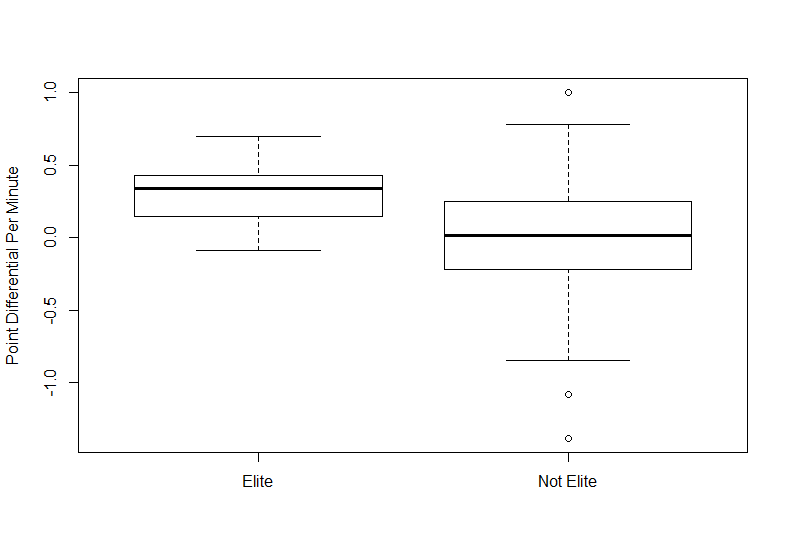}
\caption{Boxplots of lineup point differential per minute (PMM) by ANC-predicted label.}
\label{graph:boxplots}
\end{figure}

Additionally, we can compare the known PMM for lineups predicted to be \texttt{elite} against those that are not.  Figure \ref{graph:boxplots} provides comparative boxplots for the average point differential per minute of lineups predicted (or not) to be \texttt{elite}.  The mean PMM of predicted \texttt{elite} lineups is $+0.30$, which is statistically higher than the mean PMM of predicted \texttt{not elite} lineups of $-0.00049$ ($p = 0.00013$).  Thus, while the ANC does not predict PMM directly, it does a good job partitioning the teams into an \texttt{elite} group predicted to have a positive PMM and a \texttt{not elite} group predicted to have a negative PMM.    It is worth noting that there is overlap in the distributions, and the ANC misses some very good lineups (e.g., Toronto Raptors’ Delon Wright, DeMar DeRozan, Fred Van Vleet, Jakob Poeltl, and Pascal Siakam, which had 40.0 minutes of playing time and a PMM of 1.00.).  Nonetheless, the distribution of PMM of \texttt{elite} lineups is generally higher than the distribution of PMM of \texttt{not elite} lineups, and the two lineups having negative PMM that were misclassifed by the ANC to be \texttt{elite} have only moderately negative PMMs.  This points to the ANC having good judgment about elite lineups.

One might also wonder whether the complete set of five-player order statistics is required by the ANC to achieve high precision.  Analysis of a simpler model, using only the first order statistics (i.e., the minimum of each individual player statistic on the lineup), is presented in Appendix \ref{app:dummymodel}.  The simpler model achieves a testing precision of only 75\% compared to the ANC's testing precision of 86.7\%.

\subsection{Interpretation of the Subclassifiers}
\label{ss:interpret}

After training the ANC using the optimized parameter values given in Table \ref{tab:tunedparams}, we now try to interpret the classifications given by each subclassifier to understand which variables are most important in determining lineup performance.  The $k$-nearest neighbor and support vector machine are not particularly interpretable, so we focus primarily on the other five subclassifiers.

By far, the most consistently important predictor of elite classification over the five interpretable subclassifiers is the smallest plus-minus per minute of the five players on the lineup, PMM$_{(1)}$.  This is not surprising because 
if the lineup’s worst plus-minus is high, then all five players’ plus-minuses must be high, indicating a high level of play across any lineups in which the players took part.  

In the decision tree, PMM$_{(1)}$ is the only branch of the tree that remains after pruning.  Elite classification requires standardized PMM$_{(1)}  > 0.32$.

Table \ref{tab:logreg} gives the estimated coefficients of the 24 standardized predictors identified by backwards stepwise logistic regression.  The three most important predictors in the model, both in terms of coefficient magnitude and statistical significance, are PMM$_{(1)}$, the lowest plus-minus per minute among the five players, FG3A$_{(5)}$, the highest rate of 3-point field goals attempted per minute among the five players, and SCREENASSISTS$_{(3)}$, the median rate per minute of screens that led to baskets.  Additionally, we notice that a variety of offensive and defensive statistics, including hustle statistics, are selected by the model to predict lineup performance.

\begin{table}
\caption{Estimated coefficients of logistic regression model. Significance codes:   ***:0.001, **:0.01, *:0.05, .:0.1}
\label{tab:logreg}
\begin{tabular}{lll}

Predictor & Estimate & $p$-value \\  
\hline
(Intercept)   &         0.12970  &  0.08068 .  \\
FGM$_{(3)}$        &          0.17638 &   0.08454 .  \\
FGA$_{(4)}$        &         -0.16022  &   0.10885    \\
FG3M$_{(5)}$       &          0.43471  &  0.04551 *  \\
FG3A$_{(5)}$       &         -0.60677  &  0.00666 ** \\
FG3PCT$_{(2)}$    &          0.22172   &  0.01398 *  \\
FG3PCT$_{(3)}$    &         -0.24260  & 0.01370 *  \\
FTA$_{(4)}$        &          0.21057     & 0.02255 *  \\
FTPCT$_{(1)}$     &          0.19298   &  0.02183 *  \\
FTPCT$_{(3)}$     &         -0.22849   &  0.03709 *  \\
FTPCT$_{(4)}$     &          0.27984 &  0.02016 *  \\
FTPCT$_{(5)}$     &          0.20627    &  0.04374 *  \\
DREB$_{(2)}$       &          0.17627   &  0.03486 *  \\
TOV$_{(4)}$        &         -0.15526    & 0.06852 .  \\
BLK$_{(5)}$        &          0.13514    &  0.10880    \\
PF$_{(5)}$         &          0.20912  &  0.01139 *  \\
PFD$_{(1)}$        &         -0.19257   &  0.01774 *  \\
PMM$_{(1)}$ &          0.85462    & < 2e-16 ***\\
CONTESTEDSHOTS$_{(4)}$ &     -0.19241  &  0.04944 *  \\
CONTESTEDSHOTS2PT$_{(1)}$ &  0.19000   & 0.04680 *  \\
CONTESTEDSHOTS3PT$_{(5)}$ & 0.20103  & 0.03562 *  \\
SCREENASSISTS$_{(2)}$      &-0.17644  & 0.06542 .  \\
SCREENASSISTS$_{(3)}$      & 0.30631  & 0.00984 ** \\
SCREENASSISTS$_{(5)}$      & 0.23956  &  0.01057 *  \\
BOXOUTS$_{(3)}$            &-0.26324 &  0.02719 *  \\
\end{tabular}
\end{table}

\begin{figure}
\includegraphics[width =\textwidth]{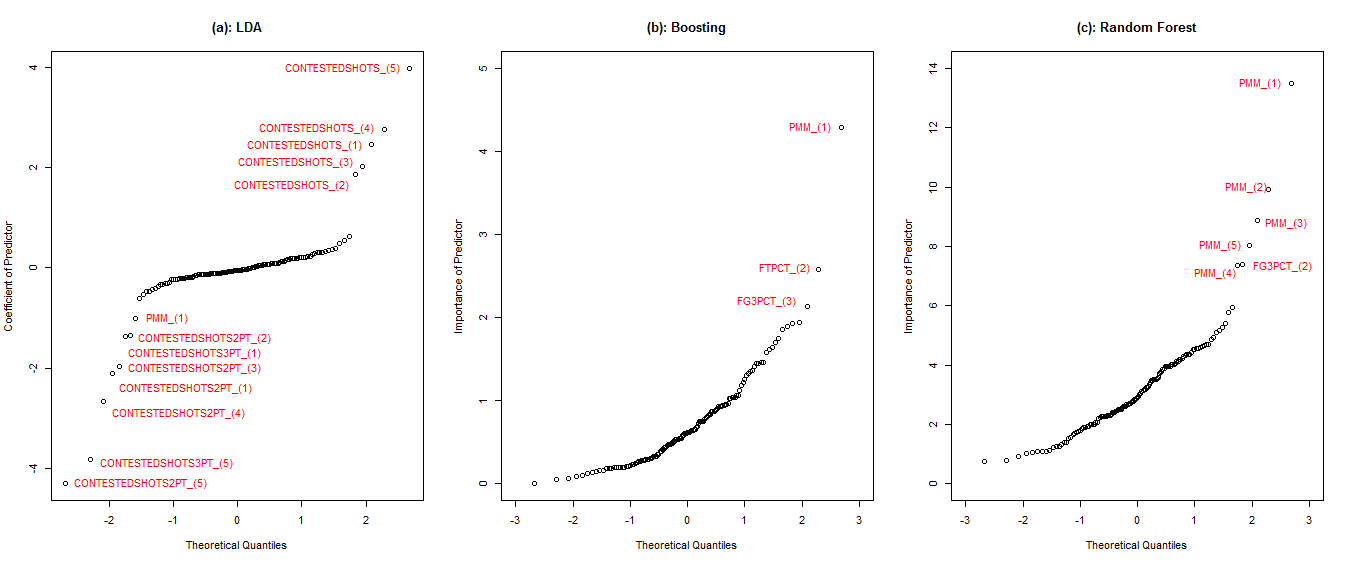}
\caption{Predictor importance for (a) linear discriminant analysis, (b) boosting, and (c) random forest subclassifiers.  Predictors listed in red are influential in that subclassifier’s determination of elite lineups.}
\label{graph:predimport}
\end{figure}

For the cases of linear discriminant analysis, boosting, and random forests, predictor importance can be assessed graphically, as shown in Figure \ref{graph:predimport}. Figure \ref{graph:predimport}(a) shows the coefficient of each standardized predictor in the linear discriminant analysis (LDA) model, plotted against quantiles of the normal distribution. The predictors in red have coefficients that are particularly large in magnitude, indicating their influence in classifying teams as \texttt{elite} or \texttt{not elite}.  In addition to PMM$_{(1)}$, statistics related to contested shots carry large importance in the LDA model.  Figure \ref{graph:predimport}(b) shows predictor importance for boosting, measured as reduction in the Gini impurity index, plotted against quantiles of the normal distribution.  Points in red have very large importance, and correspond to the predictors PMM$_{(1)}$, FTPCT$_{(2)}$ (second-lowest free throw percentage) and FG3PCT$_{(3)}$ (median 3-point field goal percentage).  Figure \ref{graph:predimport}(c) shows predictor importance for the random forest subclassifier, measured as reduction in the Gini impurity index, plotted against quantiles of the normal distribution.  Predictors with especially high importance are all five players’ plus-minuses and the second smallest 3-point field goal percentage (FG3PCT$_{(2)}$). We conclude that PMM$_{(1)}$ is consistently the most important predictor of lineup performance, followed to a lesser extent by three-point field goal percentages and other metrics.  

\subsection{Predictions for the 2018-19 Season}
\label{ss:predict}

Having a classifier that attains a high precision in identifying lineups with positive PMM on the 2017-18 season data, we now use the ANC to identify promising lineups for the thirty teams participating in the 2018-19 NBA regular season.  We refit the ANC using the combined training and testing data, standardized, from 2017-18.  We shift and scale the 2018-19 regular season data using the means and standard deviations from 2017-18 to avoid data leakage and we use the fitted ANC to predict 2018-19 lineup performance.  We can then compare our predictions against actual performance during this season.

The 2018-19 regular season data is gathered as follows. For a given team roster, we generate all possible five-player lineup combinations and their individual player order statistics for the metrics used in the ANC (listed in Appendix \ref{app:preds}).  We exclude any lineups wherein any of the players had less than 50 minutes of playing time in the 2017-18 regular season.  Of the 530 players listed on 2018-19 regular season rosters of the thirty NBA teams, 392 of these had sufficient playing time during the 2017-18 regular season and are included in the analysis. We can then predict which of these possible lineups will have a positive PMM.  This part of the process is interesting for two reasons.  First, for rebuilding teams, such as the 2018-19 Los Angeles Lakers, we can predict which combinations of newly traded and existing players are likely to play well together.  Second, we can compare the relative strength of NBA teams by measuring the depths of their benches in terms of number of elite lineups.  

Table \ref{tab:predelite} gives the number of predicted \texttt{elite} lineups for each team having more than zero elite predictions.  The six teams having the largest number of predicted \texttt{elite} lineups are the Golden State Warriors, the Houston Rockets, the Indiana Pacers, the Boston Celtics, the Utah Jazz, and the Charlotte Hornets.  The remaining teams each have four or fewer predicted \texttt{elite} lineups, and fourteen out of the thirty teams have zero predicted \texttt{elite} lineups.  By prioritizing precision, the ANC is a conservative predictor of elite lineups.


{\begin{table}
\caption{Number of ANC-predicted \texttt{elite} lineups for the 2018-19 NBA regular season for the 16 teams having nonzero elite predictions.}
\label{tab:predelite}
\begin{tabular}{ll}

Team & Number of Predicted Elite Lineups \\ \hline
Golden State Warriors	&127 \\
Houston Rockets	&62 \\
Indiana Pacers	&30 \\
Boston Celtics	&25 \\
Utah Jazz	&18 \\
Charlotte Hornets	&14 \\
Oklahoma City Thunder	&4 \\
Denver Nuggets	&4 \\
Portland Trail Blazers	&3 \\
New Orleans Pelicans	&3 \\
Minnesota Timberwolves	&3 \\
San Antonio Spurs	&2 \\
Miami Heat	&2 \\
Washington Wizards	&1 \\
Toronto Raptors	&1 \\
Philadelphia 76ers	&1 \\
\end{tabular}
\end{table}

\subsubsection{Comparing 2018-19 Lineup Predictions To Realized Performance}

We now compare the predictions made by the ANC to actual performance of lineups used during the 2018-19 regular season. 

We begin by examining the confusion matrix.  For any given roster, the vast majority of lineups (predicted to be \texttt{elite} or not) may never be used, and the lineups used in the 2018-19 regular season occasionally involve players who did not have enough 2017-18 playing time to be included in ANC predictions.  Thus, to construct the confusion matrix, we restrict our focus to those lineups appearing in the 2018-19 regular season data that were predicted by the ANC. On the left of Table \ref{tab:confusion201819}, we have the confusion matrix calculated on all 2018-19 regular season lineups for which an ANC prediction was obtained; on the right of Table \ref{tab:confusion201819}, we have the confusion matrix calculated on only those 2018-19 regular season lineups having at least 25 minutes of playing time for which an ANC prediction was obtained.  The purpose of this restriction is to focus on lineups for which PMM is well-estimated.  In the unrestricted case, we see that 67.4\% of those lineups predicted to be \texttt{elite} by the ANC experienced a positive PMM during the 2018-19 regular season, compared to a 58.1\% overall prevalence of elite lineups.  When we restrict our focus to those lineups having at least 25 minutes of playing time, the precision increases to 76.9\% compared to a prevalence of 62.1\%.  We conclude that using 2017-18 individual player order statistics in the ANC to predict 2018-19 lineup performance yields high-precision, if somewhat conservative, predictions of lineups achieving positive PMM.

\begin{table}
\caption{Confusion matrix for the tuned ANC applied to the 2018-19 regular season data. The unrestricted case on the left includes all 2018-19 regular season lineups having ANC predictions.  Of the 46 lineups predicted to be \texttt{elite}, 31 have a true label of \texttt{elite}, corresponding to a precision of 67.4\%. The restricted case on the right includes only those ANC-predicted lineups having at least 25 minutes of playing time during the 2018-19 regular season. Of the 26 lineups predicted to be \texttt{elite}, 20 have a true label of \texttt{elite}, corresponding to a precision of 76.9\%.}
\label{tab:confusion201819}
\begin{tabular}{llll || lll}

& \multicolumn{3}{c||}{Unrestricted} & \multicolumn{3}{c}{Restricted} \\ \hline
 && \multicolumn{2}{c||}{Predicted Class} &  &  \multicolumn{2}{c}{Predicted Class} \\
& & Elite &	Not Elite & &  Elite &	Not Elite \\ 
\multirow{2}{*}{\rotatebox[origin=c]{90}{\parbox[c]{1cm}{\centering True Class}}}  &  Elite &	31 &	397  &  Elite &	20 &	201  \\
&	Not Elite &	15	& 294 &   	Not Elite &	6	& 129 \\
\end{tabular}
\end{table}

The 26 lineups predicted to be elite by the ANC in the restricted case are listed in Tables \ref{tab:acquisitions_pos} (positive 2018-19 PMM) and \ref{tab:acquisitions_neg} (non-positive 2018-19 PMM), along with their 2018-19 lineup playing time.  Of the 20 lineups that the ANC predicted to be elite and ultimately were elite in 2018-19, further investigation reveals that six of these involved players that were newly acquired between the 2017-18 and 2018-19 seasons, as noted in the final column Table \ref{tab:acquisitions_pos}.

\begin{table}
\caption{2018-19 lineups predicted to be elite by the ANC classifier that had positive PMM (restricted to those lineups having at least 25 minutes of playing time, as in Table \ref{tab:confusion201819}.)  Also noted are those lineups involving newly acquired players that had not played for the team during the 2017-18 season.}
\label{tab:acquisitions_pos}
\begin{tabular}{p{0.14\textwidth} p{0.33\textwidth}p{0.06 \textwidth}p{0.05\textwidth} p{0.39\textwidth}}

Team	&	Lineup	&	Minutes Played	&	PMM	&	Note	\\ \hline
Boston Celtics	&	A. Horford, K. Irving, M. Smart, J. Brown, J. Tatum	&	56	&	0.3	&		\\ \hline
\multirow{3}{*}{\parbox{0.15\textwidth}{Charlotte Hornets}}	&	M. Williams, N.  Batum, K. Walker, J. Lamb, C. Zeller	&	593	&	0.16	&		\\
	&	M. Williams, N.  Batum, K. Walker, J. Lamb, W. Hernangomez	&	34	&	0.03	&	Willy Hernangomez played for the NY Knicks in 2017-18. 
 \\ \hline
\multirow{3}{*}{\parbox{0.15\textwidth}{Golden State Warriors}}	&	K. Durant, S. Curry, K. Thompson, D. Green, K. Looney	&	313	&	0.39	&		\\
	&	K. Durant, S. Curry, D. Cousins, K. Thompson, D. Green	&	268	&	0.29	&	DeMarcus Cousins played for the New Orleans Pelicans in 2017-18. 
 \\
	&	A. Iguodala, K. Durant, S. Curry, K. Thompson, D. Green	&	178	&	0.69	&		\\
	&	A. Iguodala, K. Durant, S. Curry, K. Thompson, K. Looney	&	141	&	0.17	&		\\
	&	A. Iguodala, K. Durant, S. Curry, K. Thompson, J. Bell	&	36	&	0.73	&		\\
	&	A. Iguodala, S. Curry, D. Cousins, K. Thompson, D. Green	&	29	&	0.77	&	DeMarcus Cousins played for the New Orleans Pelicans in 2017-18.	\\
	&	A. Iguodala, K. Durant, S. Curry, D. Green, K. Looney	&	25	&	0.8	&		\\ \hline
\multirow{3}{*}{\parbox{0.15\textwidth}{Houston Rockets}}	&	C. Paul, P. Tucker, E. Gordon, J. Harden, C. Capela	&	420	&	0.15	&		\\
	&	C. Paul, P. Tucker, J. Harden, A. Rivers, C. Capela	&	30	&	0.07	&	Austin Rivers played for the LA Clippers in 2017-18. 
 \\ \hline
\multirow{3}{*}{\parbox{0.15\textwidth}{Indianapolis Pacers}}	&	T. Young, D. Collison, B. Bogdanovic, V. Oladipo, M. Turner	&	555	&	0.1	&		\\
	&	T. Young, D. Collison, B. Bogdanovic, V. Oladipo, D. Sabonis	&	133	&	0.13	&		\\
	&	T. Young, C. Joseph, B. Bogdanovic, V. Oladipo, M. Turner	&	29	&	0.03	&		\\
	&	D. Collison, B. Bogdanovic, V. Oladipo, M. Turner, D. Sabonis	&	26	&	0.39	&		\\ \hline
Minnesota Timberwolves	&	T. Gibson, R. Covington, A. Wiggins, T. Jones, K. Towns	&	77	&	0.32	&	Robert Covington played  for the Philadelphia 76ers in 2017-18. 
\\ \hline
Oklahoma City Thunder	&	R. Westbrook, P. George, S. Adams, A. Abrines, J. Grant	&	88	&	0.21	&		\\ \hline
\multirow{3}{*}{\parbox{0.15\textwidth}{Utah Jazz}}	&	D. Favors, R. Gobert, J. Ingles, R. O'Neale, D. Mitchell	&	107	&	0.18	&		\\
	&	K. Korver, T. Sefolosha, R. Rubio, R. Gobert, D. Mitchell	&	26	&	0.7	&	Kyle Korver played for the Cleveland Cavaliers in 2017-18. 
 \\
\end{tabular}
\end{table}








\begin{table}
\caption{2018-19 lineups predicted to be elite by the ANC classifier that had non-positive PMM (restricted to those lineups having at least 25 minutes of playing time, as in Table \ref{tab:confusion201819}.)   Also noted are those lineups involving newly acquired players that had not played for the team during the 2017-18 season.}
\label{tab:acquisitions_neg}
\begin{tabular}{p{0.14\textwidth} p{0.33\textwidth}p{0.06 \textwidth}p{0.06\textwidth} p{0.39\textwidth}}

Team	&	Lineup	&	Minutes Played	&	PMM	&	Note	\\ \hline
Boston Celtics	&	A. Horford, K. Irving, A. Baynes, J. Brown, J. Tatum	&	25	&	0	&		\\ \hline
\multirow{3}{*}{\parbox{0.15\textwidth}{Golden State Warriors}}	&	K. Durant, S. Curry, J. Jerebko, K. Thompson, D. Green	&	45	&	-0.22	& Jonas Jerebko	played for the Utah Jazz in 2017-18.	\\
	&	K. Durant, S. Curry, K. Thompson, D. Green, J. Bell	&	26	&	-0.38	&		\\ \hline
\multirow{3}{*}{\parbox{0.15\textwidth}{Houston Rockets}}	&	G. Green, P. Tucker, E. Gordon, J. Harden, C. Capela	&	66	&	-0.32	&		\\
	&	C. Anthony, C. Paul, P. Tucker, E. Gordon, C. Capela	&	45	&	-0.11	&	Carmelo Anthony	played for the Oklahoma City Thunder in 2017-18.\\ \hline
Indianapolis Pacers	&	T. Young, T. Evans, D. Collison, B. Bogdanovic, D. Sabonis	&	28	&	-0.43	&		Tyreke Evans played for the Memphis Grizzlies in 2017-18.\\ \hline
\end{tabular}
\end{table}

We can also compare the realized 2018-19 lineup PMM between lineups predicted to be \texttt{elite} by the ANC to those predicted to be \texttt{not elite}.  Figure \ref{graph:boxplot2018predelite} gives boxplots of observed PMM during the 2018-19 regular season of lineups for which an ANC prediction was given.  Figure \ref{graph:boxplot2018predelite}(a) shows this in the unrestricted case, while Figure \ref{graph:boxplot2018predelite}(b) restricts to lineups that had at least 25 minutes of playing time in the 2018-19 regular season.  In both cases, we see that the PMMs of lineups predicted by the ANC to be \texttt{elite} exhibit lower variance than those predicted to be \texttt{not elite}.  Moreover, when we restrict consideration to those lineups having at least 25 minutes of playing time, for which the PMM is more precisely estimated, we see that the distribution of PMM for those predicted to be \texttt{elite} lies nearly entirely in the positive range.  A one-sided test of the means reveals that the mean PMM for those lineups predicted to be \texttt{elite} is likely higher than the mean PMM for those lineups predicted to be \texttt{not elite} ($p = 0.090$).

\begin{figure}
\includegraphics[width =\textwidth]{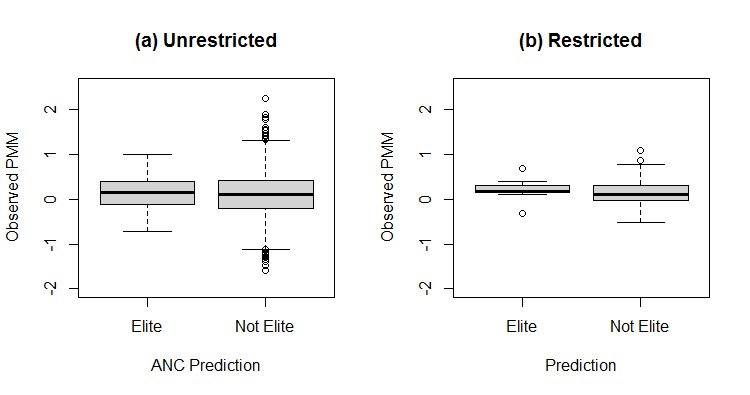}
\caption{Boxplots of 2018-19 lineup point differential per minute (PMM) by ANC-predicted label.  (a) Unrestricted case.  (b) 2018-19 lineups are restricted to those having at least 25 minutes of playing time.}
\label{graph:boxplot2018predelite}
\end{figure}

Next we explore the relationship between ANC predictions and overall team bench strength and performance.  Figure \ref{graph:PredActElite} plots, for each of the thirty NBA teams, the number of lineups used in the 2018-19 regular season that had a positive PMM versus the number of  lineups the ANC predicted would be \texttt{elite} for that team.  Figure \ref{graph:PredActElite}(a) does this for the unrestricted case, while Figure \ref{graph:PredActElite}(b) restricts the counts on the $y$-axis to those lineups having at least 25 minutes of playing time in 2018-19.  We first note that of the six teams predicted to have a relatively large number of \texttt{elite} lineups (GSW, HOU, IND, BOS, UTA, and CHA), Boston and the Golden State Warriors, and to a lesser extent, Houston, also have a relatively large number of lineups achieving positive PMM during the 2018-19 season in the unrestricted case.  When we restrict our focus to those lineups having at least 25 minutes of playing time, for which the PMM is more precisely estimated, we see in Figure \ref{graph:PredActElite}(b) that Boston, and to a lesser extent Indiana and Golden State, have a robust bench of elite lineups.  

\begin{figure}
\includegraphics[width =\textwidth]{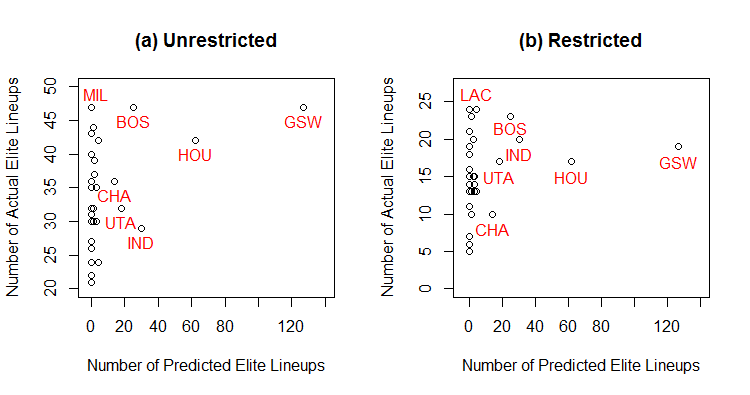}
\caption{Number of lineups on each team achieving a positive PMM during the 2018-19 regular season versus the number of lineups predicted by the ANC to be \texttt{elite} on each team. (a) Unrestricted case.  (b) 2018-19 lineups are restricted to those having at least 25 minutes of playing time.}
\label{graph:PredActElite}
\end{figure}

It is worth noting that the ANC misses some teams that ended up having many lineups with positive PMMs.  For example, in Figure \ref{graph:PredActElite}(a) we see that none of the Milwaukee Bucks' lineups were predicted to be \texttt{elite} by the ANC, and yet 47 of its lineups achieved positive PMM, 19 of which experienced at least 25 minutes of playing time; this performance is on par with that of GSW.   For the case of Milwaukee, 2018-19 turned out to be an unprecedented season with the hiring of Head Coach Mike Budenholzer. The team achieved its best regular season record in several decades, and the best regular season record in the NBA overall, won the Central Division and reached the Eastern Conference Finals \cite{BucksWiki}.  At the end of that season, Coach Budenholzer was selected to coach the East team in the 2019 NBA All-Star Game, was named NBA Coach of the Year, and was awarded by his peers the National Basketball Association's Coach of the Year Award \cite{BudenholzerWiki}.  We speculate that while the ANC does a good job capturing general lineup ability, it is unable to account for changes to coaching staff and style.

Likewise, Figure \ref{graph:PredActElite}(b) shows that the team having the highest number of lineups with positive PMM when restricted for 25 minutes playing time during 2018-19 is the Los Angeles Clippers, for which the ANC also predicts zero \texttt{elite} lineups.  In this case, 59 of the 66 lineups used by the team during the 2018-19 season involved players lacking sufficient individual playing time in 2017-18 to be considered by the ANC.  Restricting consideration to the 36 lineups that had at least 25 minutes of lineup playing time in 2018-19, 34 involved players with insufficient individual playing time in 2017-18 to be included in ANC predictions. Teams with a large number of rookie players, players who moved up from the G league, and players lacking playing time in the previous season will not obtain many predictions from the ANC.




\subsubsection{Other aspects of ANC performance}

We now discuss how ANC performance relates to player position and team pace.

\textbf{Balance of Positions.}  The ANC considers only player order statistics in its predictions of lineup performance and ignores other player information such as position.  A natural question, then, is whether the lineups predicted to be \texttt{elite} by the ANC exhibit a balance of positions.  Using player position information \cite{BBallRefPosition, BBallRefPosition18}, we define the positions to be Center, Power Forward, Point Guard (either a dedicated point guard or a point guard / shooting guard combination player), Shooting Guard (either a dedicated shooting guard, point guard / shooting guard combination player, or small forward / shooting guard combination player), and Small Forward (either a dedicated small forward, or a small forward / shooting guard combination player). We tally the number of distinct positions reflected in a given lineup.  Those lineups having more distinct positions are more \textit{balanced} than those having fewer distinct positions. Figure \ref{graph:histposition} compares the distribution of distinct positions between actual lineups, and those predicted \texttt{elite} or \texttt{not elite} by the ANC.  We see that ANC-predicted-\texttt{elite} lineups have more lineups exhibiting four or five  distinct positions than those predicted to be \texttt{not elite}.  Lineups that saw actual playtime in 2018-19 were slightly more balanced overall than those predicted \texttt{elite} by the ANC, but even 13.4\% of actual lineups used three or fewer distinct positions, among which many had positive PMM. This is indicative of a general trend in the NBA to move away from rigid positional play \cite{GuardianPositionless}.  We conclude that the ANC-predicted-\texttt{elite} lineups exhibit sufficient positional balance.

\begin{figure}
\includegraphics[width =0.8\textwidth]{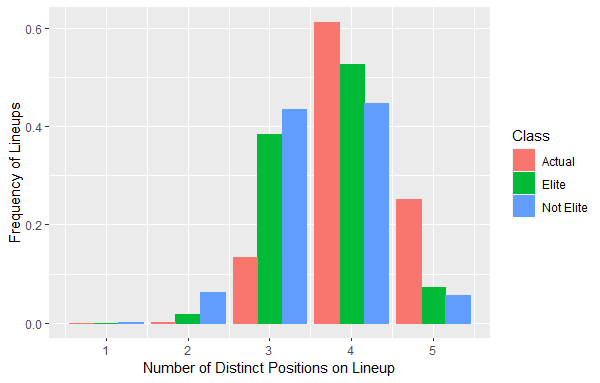}
\caption{Distribution of number of distinct positions exhibited on actual lineups, versus lineups predicted to be \texttt{elite} or \texttt{not elite} by the ANC.}
\label{graph:histposition}
\end{figure} 

Additionally, the order statistics used by the ANC capture sufficient information about positional roles such that the ANC is unlikely to recommend a lineup of five players of the same position.  To demonstrate this, we generated fictitious rosters of 2018-19 players all holding the same position; we limited the analysis to pure positions, omitting hybrid positions such as point guard / shooting guard combination.  As a proxy for player quality, we focused on players having at least 2100 minutes of playing time during the 2017-18 season (this corresponds roughly to the top quartile of playing time for each position).  For each position, we sampled 200 lineups uniformly at random from the possible 5-player lineups and used the ANC to predict whether or not that lineup would have a positive PMM.  As summarized in Table \ref{tab:posfakerosters}, there was not a single lineup predicted to be \texttt{elite} in this manner.  Thus, the order statistics used by the ANC appear to be capturing the contributions of different positions on the court.

\begin{table}
\caption{ANC predictions on lineups comprised of a single position for 2018-19 players having at least 2100 minutes of 2017-18 playing time.}
\label{tab:posfakerosters}
\begin{tabular}{llll}

Position & Players & Lineups Sampled &  Predicted \texttt{elite} \\  
\hline
Shooting Guard & 22 & 200 & 0 \\
Power Forward & 14 & 200 & 0 \\
Center & 11 & 200 & 0 \\
Point Guard & 15 & 200 & 0 \\
Small Forward & 19 & 200 & 0 \\
\end{tabular}
\end{table}

\textbf{Team Pace.}  The ANC relies on per-minute player statistics rather than per-possession statistics.  Because teams are known to vary in their pace, a natural question is whether the ANC predicts more \texttt{elite} lineups for fast-paced teams than slow-paced teams.  Figure \ref{graph:pace} shows the number of \texttt{elite}-predicted lineups for each team versus the team's pace, as reported by NBA.com \cite{NBApace}.  We observe no trend between pace and the propensity of the ANC to label lineups as \texttt{elite}. We leave for future work the incorporation of team pace into the ANC when predicting the performance of lineups having combinations of players from different teams.

\begin{figure}
\includegraphics[width =0.8\textwidth]{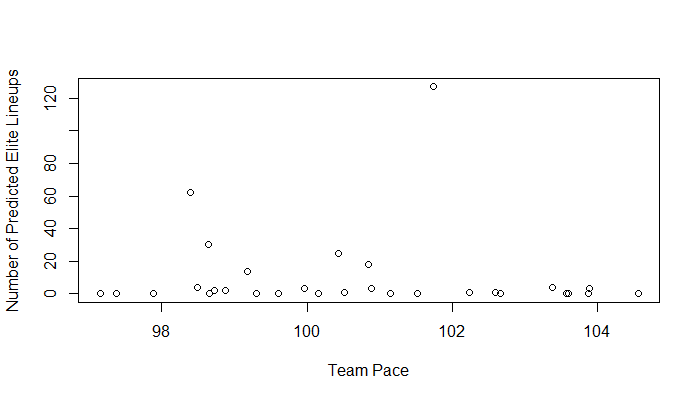}
\caption{Number of lineups on each team predicted by the ANC to be \texttt{elite} versus team pace.}
\label{graph:pace}
\end{figure}

\subsubsection{Case Study of Golden State Warriors and Los Angeles Lakers}

To shed more light on the ANC predictions for the 2018-19 regular season, we use as a case study the Golden State Warriors (NBA champions in the two preceding years), and the Los Angeles Lakers, whose acquisition of LeBron James during the 2018 off-season led many to hope at the time that this once powerful team would be staging a come-back.  

The predictions from our ANC model are summarized in Table \ref{tab:2018preds}.  The Los Angeles Lakers had fifteen 
players on their roster during this season with sufficient 2017-18 individual playing time to be included in ANC predictions, yielding 3003  
possible five-person lineups.  Of these, 
not a single lineup is predicted to be \texttt{elite}.
The Golden State Warriors also had fifteen players with sufficient 2017-18 individual playing time on their roster in 2018-19, yielding 3003 possible five-person lineups.  In contrast to the Lakers, 
127 of these lineups are predicted to be \texttt{elite}.   

To understand why the Lakers have no lineups predicted to be \texttt{elite}, we examine the individual player PMM for both the Lakers and Warriors rosters in Table \ref{tab:LakWarr} from the 2017-18 season that was used to train the model.  We see that the Lakers had several players on the roster with negative PMM.  Given the importance assigned by the subclassifiers to PMM$_{(1)}$, this sheds some insight into why the Lakers are not predicted by the ANC to have any \texttt{elite} lineup.  The lack of predicted \texttt{elite} lineups is also consistent with the team's ultimately poor performance during that season.

Entering the 2018-19 season as two-time defending champions, the Golden State Warriors, unsurprisingly, have a large proportion of total lineup combinations classified as \texttt{elite}. A sample of the 
127 predicted \texttt{elite} Warriors lineups is given in Table \ref{tab:2018preds}; it illustrates a range of lineups that the coaching staff could rely upon to optimize the use of its players whenever the ``Hamptons Five'' (Stephen Curry, Klay Thompson, Andre Iguodala, Kevin Durant, Draymond Green) were not all on the court.  Additionally, the sheer number of \texttt{elite} lineups Golden State offers is an indicator of the depth of the team and is consistent with its record at the time. 

\begin{table}
\caption{Sample \texttt{elite} lineups predicted by the ANC for the 2018-19 rosters of the Los Angeles Lakers and the Golden State Warriors.}
\label{tab:2018preds}
\begin{tabular}{p{0.25\textwidth} p{0.13\textwidth} p{0.15\textwidth} p{0.47\textwidth}}

Team &	Number of Possible Lineups	 & Number of Predicted Elite Lineups & 	Sample Elite Lineups \\ \hline
Los Angeles Lakers &	3003 &	0	 &  \begin{itemize}
    \item None
\end{itemize}\\ \hline
Golden State Warriors &	3003	& 127 & \begin{itemize} \item	D. Cousins, S. Curry, K. Durant, D. Green, A. Iguodala 

\item S. Curry, K. Durant, J. Jerebko, K. Looney, K. Thompson  

\item K. Durant, D. Green, A. Iguodala, K. Looney, K. Thompson 

\item S. Curry, J. Bell, K. Durant, K. Looney, K. Thompson

\item D. Cousins, A. Iguodala, K. Looney, S. Livingston, K. Thompson





\end{itemize} \\
\end{tabular}
\end{table}

\begin{table}
\caption{Individual player plus-minus per minute (PMM) during the 2017-18 regular season for members of the Los Angeles Lakers and Golden State Warriors 2018-19 rosters.  Excluded from the table are LAL players Isaac Bonga, Jemerrio Jones, Scott Machado, Sviatoslav Mykhailiuk, Moritz Wagner, and Johnathan Williams, and GSW players Jacob Evans and Marcus Derrickson, who did not have enough data from the 2017-18 NBA regular season; and LAL player Ivica Zubac, who was traded mid-season to the Los Angeles Clippers and is listed by NBA.com on that team's roster.}
\label{tab:LakWarr}
\begin{tabular}{ll|ll}

\multicolumn{2}{c|}{Los Angeles Lakers} &	\multicolumn{2}{c}{Golden State Warriors} \\ \hline
Player	& Individual PMM	& Player& 	Individual PMM \\ \hline
Lonzo Ball &	-0.01 &	Jordan Bell	& 0.20 \\
Michael Beasley	& -0.08	 & Andrew Bogut & -0.13\\
Reggie Bullock & 0.01 & Quinn Cook & 	-0.05  \\
Kentavious Caldwell Pope &	-0.02& Demarcus Cousins &	0.05	 \\
Alex Caruso &	-0.01 &	Stephen Curry	& 0.30\\
Tyson Chandler &	-0.21  &	Kevin Durant &	0.15 \\
Josh Hart &	-0.05 &	Draymond Green &	0.14 \\
Andre Ingram & 0.31 & Andre Iguodala	& 0.17\\
Brandon Ingram	& -0.06  & Jonas Jerebko & 0.05 \\
Lebron James &	0.03 & Damian Jones & 0.00	 \\
Kyle Kuzma	& -0.05	& Damion Lee & -0.06   \\
Mike Muscala & -0.07 & Shaun Livingston	& 0.10 \\
JaVale McGee	& 0.06	&	 Kevon Looney	& 0.13  \\
Rajon Rondo	& 0.01 & Alfonzo McKinnie & -0.22\\
Lance Stephenson &	-0.06 & Klay Thompson	& 0.16	 \\
\end{tabular}
\end{table}


We now compare the predictions of the ANC based on 2017-18 performance to realized performance during the 2018-19 regular season, for the Los Angeles Lakers and the Golden State Warriors.

Table \ref{tab:conf201819} gives, for the Lakers and Warriors respectively, the confusion matrix showing numbers of lineups predicted \texttt{elite} or \texttt{not elite} versus their true performance (positive PMM versus nonpositive PMM), for those lineups that had at least 25 minutes of playing time during the 2018-19 regular season \cite{nbacomPMM}. We see that of the nine GSW lineups predicted to be \texttt{elite} by the ANC based on 2017-18 data, seven of these ultimately had a positive PMM in 2018-19, corresponding to a precision of 77.8\%.  (The precision cannot be calculated for the Lakers lineups because no lineups were predicted to be \texttt{elite} by the ANC.)

\begin{table}[H]
\caption{Confusion matrices for Los Angeles Lakers and Golden State Warriors, respectively, ANC predictions based on 2017-18 data compared to 2018-19 actuals, for lineups that had at least 25 minutes of playing time during the 2018-19 regular season \textit{and} whose players had at least 50 minutes of playing time during the 2017-18 regular season.}
\label{tab:conf201819}
\begin{tabular}{llll|llll}

\multicolumn{4}{c|}{Los Angeles Lakers} & \multicolumn{4}{c}{Golden State Warriors} \\ \hline
&& \multicolumn{2}{c|}{Predicted Class} & && \multicolumn{2}{c}{Predicted Class}	\\
&& Elite & Not Elite & && Elite & Not Elite \\
\multirow{2}{*}{\rotatebox[origin=c]{90}{\parbox[c]{1cm}{\centering True Class}}}  &  Elite   &      0    &     12 &\multirow{2}{*}{\rotatebox[origin=c]{90}{\parbox[c]{1cm}{\centering True Class}}}  & Elite & 7 & 12 \\
&  Not Elite  &  0      &   9 && Not Elite & 2 & 4 \\
\end{tabular}
\end{table}

Tables \ref{tab:actualPMMLAL} and \ref{tab:actualPMMGSW} in Appendix \ref{app:LALGSW} give realized PMM for all five-person lineups for the Lakers and Warriors, respectively, that had at least 25 minutes of playing time during the 2018-19 regular season, along with the ANC prediction.  `$-$' denotes lineups for which no ANC prediction is given.   
We confirm that while the ANC consistently predicts low-PMM lineups as \texttt{not elite}, it is a conservative classifier, occasionally missing lineups that actualized high PMMs the following season.  For example, the highest performing lineup of the two teams, GSW's McKinnie, Green, Looney, Livingston and Curry, is predicted by the ANC to be \texttt{not elite}.  This \texttt{not elite} prediction could be due to Alfonzo McKinnie's negative individual PMM in 2017-18, and the relatively high actual PMM could be due to imprecision caused by this lineup having only 28 minutes of playing time during the 2018-19 season.  


\section{Conclusion}
\label{s:conclusion}


We have developed an all-or-nothing classifier that predicts, based on individual player order statistics, the performance of a five-person basketball lineup.  By tuning the classifier to achieve high precision, the lineups predicted \texttt{elite} by the ANC achieve a higher average plus-minus per minute of playing time than teams predicted to be \texttt{not elite}, both in same-season predictions on the withheld testing data set and in next-season predictions.  We also showed that teams having a large number of predicted-\texttt{elite} lineups tend to have a large number of lineups that achieve high PMM, demonstrating that the ANC can be an indicator of bench depth.  By incorporating both offensive and defensive individual player order statistics, the ANC captures positional play and recommends lineups that are consistent in positional balance with those used in actual play.  

While the lineups classified as \texttt{elite} by the ANC do have a statistically higher PMM than lineups classified as \texttt{not elite}, the ANC occasionally misses very high-performing lineups.  One possibility for future work is to tune the PMM threshold at which a lineup is labeled as \texttt{elite}.  Our current threshold is zero, but we might be interested in identifying and characterizing lineups whose PMM is some amount higher than that; the classifier might work better identifying more extreme cases. Additionally, evidence suggests the ANC loses its predictive power for teams that undergo changes to coaching staff and style and for teams having a large number of novice players.  Future work could examine whether past-year data from other leagues (e.g. the G league or NCAA) can be used to make predictions those players without sufficient NBA history.

Nonetheless, because the ANC prioritizes precision, lineups predicted by the ANC to be \texttt{elite} often actualize positive PMM.  We conclude that this classifier can be included among decision-support tools to inform player acquisitions and substitutions.

\textbf{Acknowledgement}
This material is based upon work supported by the National Science Foundation under Grant No. DMS-1757952. Any opinions, findings, and conclusions or recommendations expressed in this material are those of the author and do not necessarily reflect the views of the National Science Foundation.  The authors would also like to acknowledge financial support from Harvey Mudd College.  The authors thank Isys Johnson,  Lucius Bynum, and Robert Gonzalez for their contributions to earlier phases of this work and to the code base, portions of which were adapted and used in this paper. 
 Lastly, the authors thank the anonymous reviewers and editors whose feedback greatly improved the analysis.   

\bibliographystyle{chicago}
\bibliography{ref.bib}

\appendix
\section{Individual Player Statistics Used as Predictors in ANC}
\label{app:preds}
\begin{table}[H]
\caption{Individual player statistics used as predictors in ANC.}
\label{tab:predictors}
\begin{tabular}{ll}

FGM & Field goals made per minute \\
FGA & Field goals attempted per minute \\
FGPCT & Field goal percentage \\
FG3M & Three-point field goals made per minute \\
FG3A & Three-point field goals attempted per minute \\
FG3PCT & Three-point field goals percentage \\
FTM & Free throws made per minute \\
FTA & Free throws attempted per minute \\
FTPCT & Free throw percentage \\
OREB & Offensive rebounds per minute \\
DREB & Defensive rebounds per minute \\
AST & Assists per minute \\
TOV & Turnovers per minute \\
STL & Steals per minute \\
BLK & Blocks per minute \\
BLKA & Blocks attempted per minute \\
PF & Personal fouls per minute \\
PTS & Points earned per minute \\
PFD & Personal fouls drawn per minute \\
PMM & Plus-Minus per minute \\
CONTESTEDSHOTS & Shots contested per minute \\
CONTESTEDSHOTS2PT & Two-point shots contested per minute \\
CONTESTEDSHOTS3PT & Three-point shots contested per minute \\
CHARGESDRAWN & Charges Drawn per minute \\
DEFLECTIONS & Passes deflected per minute \\
LOOSEBALLSRECOVERED & Loose balls recovered per minute \\
SCREENASSISTS & Screens that led to baskets per minute \\
BOXOUTS & Box outs per minute \\
\end{tabular}
\end{table}
\newpage

\section{Comparison of ANC to Simpler Model}
\label{app:dummymodel}

One might also wonder whether the complete set of five-player order statistics is required by the ANC to achieve high precision.  In this section, we analyze a simpler model that uses only the first order statistics (i.e., the lineup's minimum) of each individual player metric used by the ANC. 

We tune the simple model parameters as in Section \ref{ss:tuneresults}, using ten-fold cross-validation.  The parameter combination that lies on the efficient frontier of average precision and worst-case precision over the folds on the training data  is given in Table \ref{app:tunedparamsdummy}.  This combination achieved an average precision of 86.5\%, minimum precision of 57.1\% and average accuracy of 51.8\% on the training data.  When the performance was insensitive to a parameter value, the value was chosen to match that used in the ANC.

\begin{table}
\caption{Tuned parameter values used in simple model based on first order statistics.}
\label{app:tunedparamsdummy}
\begin{tabular}{lll}

Subclassifier &	Parameter&	Chosen Value \\ \hline
Decision Tree &	\texttt{cp} (cost complexity) &	-1 \\
& \texttt{loss} (misclassification penalty) & 1\\ \hline
Random Forest &	\texttt{c} (cutoff) &	0.7 \\ 
& \texttt{ntree} (number of trees) & 100\\ \hline
Boosting &	\texttt{mfinal} (number of trees) & 500 \\
 &\texttt{maxdepth} (depth of each tree) & 3 \\
& \texttt{cp} (cost complexity) & 0.01 \\ \hline
Support Vector Machine	& \texttt{cost} (misclassification penalty)& 0.1 \\
 & \texttt{gamma} (influence decay)	& 0.01 \\ \hline
K-Nearest Neighbors	& \texttt{k} (number of neighbors) &	5 \\ \hline
Logistic Regression	& \texttt{thresh} ($1-$probability threshold) &	0.25 \\ \hline
All-or-Nothing Classifier (ANC) &	\texttt{numVotes} (agreement required) &	7 \\
\end{tabular}
\end{table}

Having tuned the parameters, we fit the first order statistic model to the full, standardized, training set, as described earlier, and apply the trained model to the testing data.  The confusion matrix is given in Table \ref{tab:confusiondummy}.

Of twelve lineups predicted to be \texttt{elite}, nine of these have a true label of \texttt{elite}, indicating a strictly positive PMM.  The simpler model achieves a testing precision of only 75\% compared to the ANC's testing precision of 86.7\%.

\begin{table}
\caption{Confusion matrix for the simple model based on first order statistics applied to the test data set. Of the 12 lineups predicted to be \texttt{elite}, nine have a true label of \texttt{elite}, corresponding to a precision of 75.0\%.}
\label{tab:confusiondummy}
\begin{tabular}{llll}

 && \multicolumn{2}{c}{Predicted Class} \\
& & Elite &	Not Elite \\ 
\multirow{2}{*}{\rotatebox[origin=c]{90}{\parbox[c]{1cm}{\centering True Class}}}  &  Elite &	9 &	86 \\
&	Not Elite &	3	& 78 \\
\end{tabular}
\end{table}

\newpage

\section{Actual Lineup Performance for LAL and GSW Case Study}
\label{app:LALGSW}

{\small
\begin{table}[h!]
\caption{Actual lineup performance compared to ANC predictions for the Los Angeles Lakers during the 2018-19 season, for all lineups having at least 25 minutes of playing time. `$-$' denotes lineups for which no ANC prediction is given.  }
\label{tab:actualPMMLAL}
\begin{tabular}{p{0.65\textwidth} p{0.1\textwidth} p{0.1\textwidth} p{0.15\textwidth}}

\multicolumn{4}{c}{Los Angeles Lakers} \\ \hline
Lineup	&	Minutes Played & Actual PMM	&	ANC Prediction	\\ \hline
R. Rondo, K. Caldwell-Pope, B. Ingram, I. Zubac\tablefootnote{Ivica Zubac was traded to the Los Angeles Clippers and was not included in ANC predictions for the Lakers.}, J. Hart	& 25 &	0.68	& $-$	\\
L. James, B. Ingram, I. Zubac, L. Ball, K. Kuzma	&	55 & 0.36	&	$-$	\\
T. Chandler, L. James, K. Caldwell-Pope, L. Ball, K. Kuzma	&	39 & 0.36	&	Not Elite	\\
L. James, J. McGee, L. Ball, K. Kuzma, J. Hart	&	133 & 0.31	&	Not Elite	\\
L. James, R. Rondo, J. McGee, K. Caldwell-Pope, K. Kuzma	&	47 & 0.23	&	Not Elite	\\
T. Chandler, L. Stephenson, K. Caldwell-Pope, B. Ingram, J. Hart	& 37 &	0.21	&	Not Elite	\\
T. Chandler, B. Ingram, L. Ball, K. Kuzma, J. Hart	&	36 & 0.14	 &	Not Elite	\\
T. Chandler, L. James, B. Ingram, L. Ball, K. Kuzma	&	61 & 0.13	&	Not Elite	\\
L. James, R. Rondo, J. McGee, K. Caldwell-Pope, B. Ingram	& 31 &	0.13	&	Not Elite	\\
B. Ingram, I. Zubac, L. Ball, K. Kuzma, J. Hart	& 39 &	0.13	&	$-$	\\
L. James, J. McGee, K. Caldwell-Pope, L. Ball, K. Kuzma	& 34 &	0.12	&	Not Elite	\\
L. James, J. McGee, R. Bullock, B. Ingram, K. Kuzma	& 73 &	0.11	&	Not Elite	\\
T. Chandler, K. Caldwell-Pope, B. Ingram, K. Kuzma, J. Hart	& 31 &	0.10 &	Not Elite	\\
T. Chandler, K. Caldwell-Pope, B. Ingram, L. Ball, K. Kuzma	& 45 &	0.04	&	Not Elite	\\
T. Chandler, L. James, L. Ball, K. Kuzma, J. Hart	& 66 &	0.02	&	Not Elite	\\
L. James, R. Rondo, J. McGee, B. Ingram, K. Kuzma	& 43 &	0.00	&	Not Elite	\\
L. James, J. McGee, B. Ingram, L. Ball, K. Kuzma	& 234 &	0.00	&	Not Elite	\\
L. James, R. Rondo, R. Bullock, B. Ingram, K. Kuzma	& 62 &	-0.05	&	Not Elite	\\
J. McGee, K. Caldwell-Pope, M. Muscala, A. Caruso, J. Jones\tablefootnote{Jemerrio Jones did not have data from the 2017-18 NBA regular season.}	& 31 &	-0.06	&	$-$	\\
L. James, K. Caldwell-Pope, L. Ball, K. Kuzma, J. Hart	& 25 &	-0.16	&	Not Elite	\\
L. James, R. Rondo, J. McGee, R. Bullock, K. Kuzma	& 62 &	-0.21	&	Not Elite	\\
R. Rondo, K. Caldwell-Pope, B. Ingram, I. Zubac, K. Kuzma	& 29 &	-0.24	&	$-$	\\
L. James, L. Stephenson, L. Ball, K. Kuzma, J. Hart	& 31 &	-0.25	&	Not Elite	\\
R. Rondo, M. Beasley, K. Caldwell-Pope, B. Ingram, I. Zubac	& 25 &	-0.28	&	$-$	\\
J. McGee, K. Caldwell-Pope, B. Ingram, L. Ball, J. Hart	& 25 &	-0.32	&	Not Elite	\\
L. James, R. Rondo, B. Ingram, I. Zubac, K. Kuzma	& 33 &	-0.43	&	Not Elite	\\
J. McGee, B. Ingram, L. Ball, K. Kuzma, J. Hart	& 83 & 	-0.47	&	Not Elite	\\
R. Rondo, J. McGee, K. Caldwell-Pope, A. Caruso, M. Wagner\tablefootnote{Moritz Wagner did not have data from the 2017-18 NBA regular season.}	& 27 &	-1.31	&	$-$	\\
\end{tabular}
\end{table}}

{\small
\begin{table}[h!]
\caption{Actual lineup performance compared to ANC predictions for the Golden State Warriors during the 2018-19 season, for all lineups having at least 25 minutes of playing time.`$-$' denotes lineups for which no ANC prediction is given.  }
\label{tab:actualPMMGSW}
\begin{tabular}{p{0.65\textwidth} p{0.1\textwidth} p{0.1\textwidth} p{0.15\textwidth}}

\multicolumn{4}{c}{Golden State Warriors} \\ \hline
Lineup	& Minutes Played &	Actual PMM	&	ANC Prediction	\\ \hline
A. McKinnie, D. Green, K. Looney, S. Livingston, S. Curry, 	&  28&	1.01	&	Not Elite	\\
A. Iguodala, D. Green, K. Durant, K. Looney, S. Curry	&	25 &0.80	&	Elite	\\
A. Iguodala, D. Cousins, D. Green, K. Thompson, S. Curry	& 29&	0.77	&	Elite	\\
A. Iguodala, J. Bell, K. Durant, K. Thompson, S. Curry	& 36 &	0.73	&	Elite	\\
A. Iguodala, D. Green, K. Durant, K. Thompson, S. Curry	& 178 &	0.69	&	Elite	\\
A. Iguodala, K. Durant, K. Looney, K. Thompson, Q. Cook	& 35 &	0.63	&	Not Elite	\\
A. McKinnie, J. Jerebko, K. Durant, K. Looney, S. Curry	& 26 &	0.61	&	Not Elite	\\
D. Green, K. Durant, K. Looney, K. Thompson, S. Curry	& 313 &	0.39	&	Elite	\\
D. Cousins, D. Green, K. Durant, K. Thompson, S. Curry	& 268 &	0.29	&	Elite	\\
A. Bogut, D. Green, K. Durant, K. Thompson, S. Curry	& 83 &	0.27	&	Not Elite	\\
A. Iguodala, D. Cousins, D. Green, K. Thompson, S. Livingston	& 67 &	0.24	&	Not Elite	\\
D. Jones, J. Jerebko, K. Durant, K. Thompson, Q. Cook	& 29 &	0.20	&	Not Elite	\\
A. McKinnie, A. Iguodala, K. Durant, K. Looney, S. Curry	& 48 &	0.19	&	Not Elite	\\
A. Iguodala, K. Durant, K. Looney, K. Thompson, S. Curry	& 141 &	0.17	&	Elite	\\
A. Iguodala, J. Jerebko, K. Durant, K. Looney, K. Thompson	& 47 &	0.13	&	Not Elite	\\
A. McKinnie, A. Iguodala, J. Jerebko, K. Looney, S. Curry	& 27 &	0.11	&	Not Elite	\\
D. Jones, D. Green, K. Durant, K. Thompson, S. Curry	& 142 &	0.11	&	Not Elite	\\
A. Iguodala, D. Green, J. Jerebko, S. Livingston, S. Curry	& 54 &	0.07	&	Not Elite	\\
A. Iguodala, D. Cousins, K. Thompson, Q. Cook, S. Livingston & 39	&	0.05	&	Not Elite	\\
A. Iguodala, D. Green, J. Jerebko, K. Thompson, S. Livingston & 26	&	0.00	&	Not Elite	\\
D. Lee, J. Jerebko, K. Looney, K. Thompson, S. Livingston & 30	&	0.00	&	Not Elite	\\
D. Green, J. Jerebko, K. Durant, K. Thompson, S. Curry	& 45 &	-0.22	&	Elite	\\
A. Iguodala, D. Jones, K. Durant, K. Thompson, Q. Cook	& 77 &	-0.33	&	Not Elite	\\
D. Green, J. Bell, K. Durant, K. Thompson, S. Curry	& 26 &	-0.38	&	Elite	\\
A. McKinnie, D. Cousins, D. Green, K. Durant, S. Curry	& 32 &	-0.56	&	Not Elite	\\
A. McKinnie, J. Evans\tablefootnote{Jacob Evans did not have data from the 2017-18 NBA regular season.}, J. Jerebko, J. Bell, Q. Cook	& 37 &	-0.57	&	$-$	\\
\end{tabular}
\end{table}}


\end{document}